%% file: apssamp.tex
\documentclass[%
reprint,
superscriptaddress,
 amsmath,amssymb,
 aps,
floatfix,
]{revtex4-2}

\usepackage{natbib}
\usepackage{graphicx}
\usepackage{dcolumn}
\usepackage{bm}
\usepackage{colortbl}
\usepackage{color,soul}

\begin{document}


\title{Charge dynamics in the 2D/3D semiconductor heterostructure WSe$_2$/GaAs}

\author{Rafael R. Rojas-Lopez}
   \email{rrlopez@fisica.ufmg.br}
\affiliation{Zernike Institute for Advanced Materials, University of Groningen, 9747 AG Groningen, The Netherlands.}
\affiliation{Departamento de Física, Universidade Federal de Minas Gerais, 31270-901, Belo Horizonte, Brazil.}
\author{Freddie Hendriks}%
\affiliation{Zernike Institute for Advanced Materials, University of Groningen, 9747 AG Groningen, The Netherlands.}
\author{Caspar H. van der Wal}%
\affiliation{Zernike Institute for Advanced Materials, University of Groningen, 9747 AG Groningen, The Netherlands.}
\author{Paulo S. S. Guimarães}
\affiliation{Departamento de Física, Universidade Federal de Minas Gerais, 31270-901, Belo Horizonte, Brazil.}
\author{Marcos H. D. Guimarães}%
    \email{m.h.guimaraes@rug.nl}
\affiliation{Zernike Institute for Advanced Materials, University of Groningen, 9747 AG Groningen, The Netherlands.}

\begin{abstract}

Understanding the relaxation and recombination processes of excited states in two-dimensional (2D)/three-dimensional (3D) semiconductor heterojunctions is essential for developing efficient optical and (opto)electronic devices which integrate new 2D materials with more conventional 3D ones.
In this work, we unveil the carrier dynamics and charge transfer in a monolayer of WSe$_2$ on a GaAs substrate.
We use time-resolved differential reflectivity to study the charge relaxation processes involved in the junction and how they change when compared to an electrically decoupled heterostructure, WSe$_2$/hBN/GaAs.
We observe that the monolayer in direct contact with the GaAs substrate presents longer optically-excited carrier lifetimes (3.5 ns) when compared with the hBN-isolated region (1 ns), consistent with a strong reduction of radiative decay and a fast charge transfer of a single polarity.
Through low-temperature measurements, we find evidence of a type-II band alignment for this heterostructure with an exciton dissociation that accumulates electrons in the GaAs and holes in the WSe$_2$.
The type-II band alignment and fast photo-excited carrier dissociation shown here indicate that WSe$_2$/GaAs is a promising junction for new photovoltaic and other optoelectronic devices, making use of the best properties of new (2D) and conventional (3D) semiconductors.

\end{abstract}

                          
\maketitle


Transition metal dichalcogenides (TMDs) have received a lot of attention because of their atomically-thin thickness and interesting optical and electronic properties \cite{Mak2010, Splendiani2010, Mak2016}.
Their thickness confines the charges in the plane of the monolayer, resulting in strikingly different properties from their bulk counterpart \cite{Mak2016, Berkelbach2013, You2015}.
Additionally, the stacking and/or twisting of consecutive monolayers into heterostructures has been shown to give rise to new physical phenomena and makes them strong candidates for the next generation of nanodevices \cite{Wilson2021, Ciarrocchi2022}.
Their low dimensionality also makes them very sensitive to local changes, such as defects in the crystal lattice, strain, or impurities \cite{Abramson2018, Watson_2021, Jasinski_2022}.
The interaction with the environment can also modify the properties of the two-dimensional (2D) semiconductor through, for instance, the interaction with gases or substrates with different electronic properties \cite{Buscema2014, Sun2017}. 
The dielectric environment for the Coulomb interaction that gives place to excitonic phenomena in TMDs is particularly important and has been shown to be able to modulate its optical properties \cite{Raja2019}.
Therefore, we can use this as an advantage for developing new nanodevices such as gas sensors, photodetectors, and solar cells \cite{Zheng2021, Xu2016, Lin2015}. 

Gallium arsenide (GaAs) is one of the most studied semiconductors because of its applications in electronics as well as its very high electronic mobility, which allows for efficient gate-induced quantum confinement to one or two dimensions \cite{Bracker2005}.
In particular, previous studies have demonstrated that the junction of this three-dimensional (3D) semiconductor with TMDs (i.e., 2D semiconductors) is a promising junction for optoelectronic devices \cite{Xu2016, Lin2015, Sarkar2020, Jia2019, Rojas-Lopez2021}.
In order to optimize and manipulate such systems for improving the design of new (opto)electronic devices, we need to obtain a high level of understanding of the electronic properties and time-evolution of their excited states.
Nonetheless, the charge dynamics and band alignment between these materials are still largely unexplored.

\begin{figure*}[ht!]
    \centering
    \includegraphics[scale=0.84]{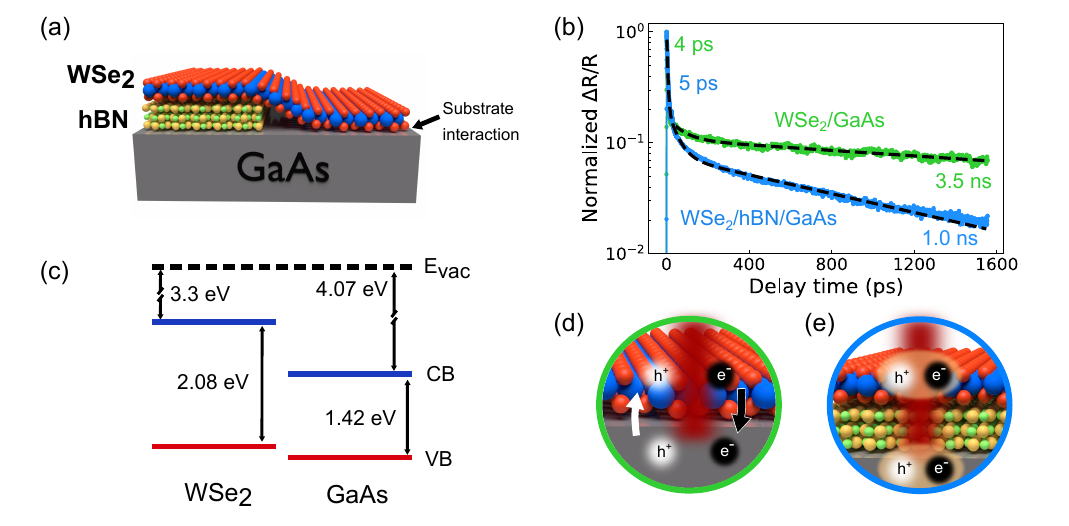}
    \caption{(a) Schematics of our sample indicating the two regions of interest WSe$_2$/hBN/GaAs, and WSe$_2$/GaAs. (b) Normalized differential reflectivity of the two regions using a laser energy for excitation in resonance with the WSe$_{2}$ exciton. (c) Estimated band alignment of the 2D/3D semiconductor heterojunction with $E_{vac}$ the vacuum energy, CB the energy of the bottom of the conduction band, and VB the top of the valence band. Presented values consider T = 300~K.  (d) Representation of the excitons generated in the WSe$_2$/GaAs region of the sample. 
    Photo-excited excitons in the monolayer and in the substrate dissociate generating an excess of electrons in the GaAs and of holes in the WSe$_2$. (e) In the WSe$_2$/hBN/GaAs region, the hBN prevents the charge transfer, allowing a more dominant role to radiative recombination processes.}
    \label{Fig1:summary}
\end{figure*}

In this work, we study the carrier dynamics in a monolayer of WSe$_2$ in contact with a GaAs substrate.
We use an optical pump-probe approach, by measuring the time-resolved differential reflectivity (TRDR) of the junction and compare it with an electrically-isolated WSe$_2$, by adding a hexagonal boron nitride (hBN) layer (Fig \ref{Fig1:summary}.a).
The WSe$_2$ monolayer in direct contact with the GaAs shows carriers that decay much slower with respect to the isolated WSe$_2$ at low temperatures.
This can be understood through a type-II band alignment that dissociates the optically-excited excitons and creates an excess of electrons in the GaAs substrate and an excess of holes in the WSe$_2$ layer.
Nonetheless, at room temperature we did not observe any important differences in the dynamics between the two regions, indicating a strong role of thermal effects on the relaxation process of photoexcited carriers.

Our samples were fabricated by mechanical exfoliation of WSe$_2$ and hBN from their bulk crystals (supplied by HQ Graphene).
The hBN flakes were exfoliated directly onto a commercial undoped (100) GaAs substrate (supplied by Wafer Technology) and the WSe$_2$ monolayers transferred on top by the viscoelastic stamp method \cite{Castellanos-Gomez2014}.
We identified WSe$_2$ monolayers by optical contrast and photoluminescence in an optical microscope.
The hBN thickness for the sample for which the results are shown here was (21$\pm$2) nm, determined by atomic force microscopy.
Time-resolved measurements were performed with a tunable Ti:Sapphire pulsed laser with a pulse width $<$ 300 fs.
We used a single-color (degenerate) pump-probe technique in a double modulation configuration as described in detail in our previous works \cite{Guimaraes2018, Rojas-Lopez2023}.
All measurements were carried out at a temperature of 70 K unless otherwise indicated.


Figure \ref{Fig1:summary}.b shows the normalized TRDR of the two regions of the sample: the direct contact (WSe$_2$/GaAs - green) and the isolated (WSe$_2$/hBN/GaAs - blue) heterostructures when excited in resonance with the exciton transition of the WSe$_2$ layer as identified by TRDR spectroscopy (see below). 
Our results are well described by a three-processes exponential decay fit, $\Delta R/R= \sum R_{0i} e^{-t/\tau _i}$, with $i$ from 1 to 3.
Such multi-exponential decay has been reported by several works in literature, but the origin of the different decay processes has been attributed to various sources, depending on the specifics of the system. 
Overall, it has been observed that radiative processes occur in no longer than a few hundred picoseconds, while non-radiative phenomena may last longer \cite{Tengfei2014, Wang2014, Qiannan2014, Tengfei2017, Ceballos2017}.

From our results, we observe a longer-lived component determined to be (3.50 $\pm$ 0.04) ns in the region of direct contact compared with (1.00 $\pm$ 0.01) ns for the isolated one.
To understand the origin of this difference, it is necessary to look into the bandgap alignment between the materials as it provides a picture of the possible charge dynamics in a junction.
For instance, previous reports observed that a junction of semiconductors with a type-I band alignment can result in a reduction of the lifetime of the material with the larger band gap when placed in such a junction \cite{wu2019, Li2020}. 
This phenomenon can be associated with an energy transfer process where, for instance, the optically generated exciton in one material transfers energy generating an exciton in the other material \cite{wang2022energy}.
On the other hand, a type-II band offset has been observed to increase the lifetime of the studied process \cite{Zereshki2019, Ceballos2017}.
In those cases, the photo-generated excitons dissociate, resulting in a charge transfer, with electrons lying in one material and holes in the other.
In light of this, our measurements point towards the existence of a type-II band offset in the WSe$_2$/GaAs heterojunction.

Simple band alignment estimations, as shown in Figure \ref{Fig1:summary}.c, further corroborate the proposed type-II band offset between monolayer WSe$_2$ and GaAs.
Here, we consider an electron affinity of 3.3 eV and an electronic bandgap of 2.08 eV for WSe$_2$, as determined in a previous work \cite{Tangi2017}. 
While the bandgap $E_g$ and electronic affinity $\chi$ of GaAs are well-established in the literature, for WSe$_2$ these values can change from one reference to another.
The sensitivity of monolayer TMDs with the electric environment and other experimental and theoretical details can lead to a variation of these values, making it challenging for an accurate determination of these properties in a generic fashion.
Nevertheless, even if differences in the exact values may arise, we can set an upper boundary for a type-II band offset. 
For this condition, the valence band maximum of WSe$_2$ has to be higher than the valence band of the GaAs:
$\chi_{WSe_2}+E_{g(WSe_2)} < 5.49$ eV.
For simplicity, here we do not take into account band bending effects due to surface states, which should be considered for a more accurate model.

This type-II band alignment implies a dissociation of photo-excited carriers with a charge transfer between the two materials. 
When the junction - monolayer and substrate - is excited, electrons at the conduction band will accumulate in the GaAs substrate, while the holes in the valence band will concentrate in the WSe$_2$ monolayer (Figure \ref{Fig1:summary}.d).
As a result, longer lifetimes of the excited states can be linked to a larger role of non-radiative scattering processes and a lack of available states in the valence (conduction) band for electrons (holes) to relax radiatively. 
In contrast, when considering the case of a WSe$_2$ isolated by hBN, charge transfer is restrained, and as a result, radiative exciton recombination is again the faster pathway for the relaxation of carriers (see Figure \ref{Fig1:summary}.e)
Therefore, our results point towards a photoexcited carrier transfer between GaAs and WSe$_2$.

In Figure \ref{Fig1:summary}b, we also observe a fast decay time ($\tau_1$) of 4 ps for the WSe$_2$ in direct contact with GaAs and 5 ps for the isolated region.
We attribute this fast process in part to a stimulated emission, related to our single-color pump-probe excitation, as well as to exciton recombination out of thermal equilibrium \cite{Tengfei2014, Wang2014, Qiannan2014, Robert2016}. 
Despite the resolution of our measurements, we cannot associate the small difference in the $\tau_1$ relaxation times as arising exclusively from the interaction with the substrate, as stress or defects in the monolayer can modify the charge dynamics within the observed difference. 
Finally, we observed an intermediate decay time ($\tau_2$) of 90 ps for the monolayer in direct contact and 50 ps in the isolated region, consistent with previous measurements of trion recombination lifetime \cite{Tengfei2017, Wang2014}.
We associate the difference in the relaxation times $\tau_2$ with an increase in the density of one type of charge carrier in the WSe$_2$ that protects the trion from fast recombination.
In particular, in our sample, two phenomena can give origin to this imbalance of carriers: the reduction of the Fermi level of the WSe$_2$ due to the formation of the joint-Fermi level of the heterojunction with the GaAs, and the dissociation and charge transfer from the photo-excited carriers \cite{Rojas-Lopez2021}. 
Moreover, the interaction of the WSe$_2$ with the GaAs can increase the dielectric disorder, which can result in an increase of recombination centers, such as defects and localized states \cite{Raja2019, Li2021}.

\begin{figure}[t!]
    \centering
    \includegraphics[width=\linewidth]{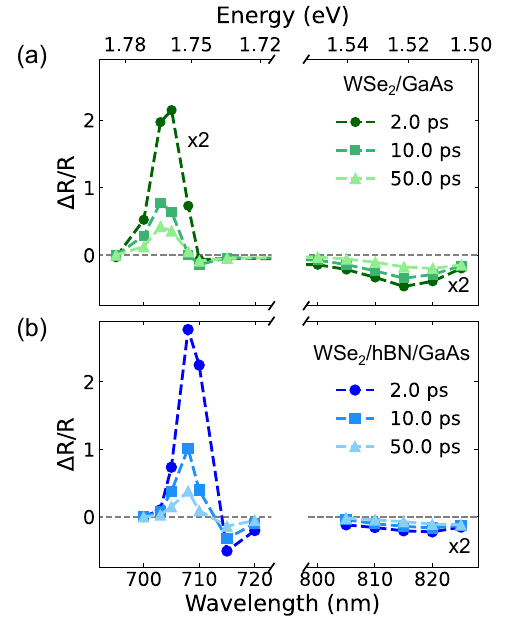}
    \caption{(a) TRDR intensity as a function of the excitation and probing wavelengths at 2 ps, 10 ps, and 50 ps pump-probe delay time in WSe$_2$/GaAs and (b) WSe$_2$/hBN/GaAs. For easier comparison the intensity values are presented as twice their real value in (a) and in the large wavelength region in (b). }
    \label{Fig2:wavelength}
\end{figure}

In order to gain further insight into the properties of our 2D/3D semiconductor junction, we study the dependence of the dynamics with the excitation wavelength.
Figure \ref{Fig2:wavelength} shows the intensity of the TRDR signal as a function of the laser wavelength at 2, 10, and 50 ps in the two regions of our system, direct contact and hBN separated.
We observe that optical resonance is different for the two regions: 705 nm for WSe$_2$/GaAs and 708 nm for the WSe$_2$/hBN/GaAs region, indicating a blue-shift on the signal of the WSe$_2$ exciton in direct contact with the GaAs.
We associate this effect with a combination of the interaction of the WSe$_2$ with a different dielectric environment and a possible effect of strain induced by the transfer onto GaAs, which should be reduced in the hBN region due to its higher smoothness and lack of dangling bonds.
We also observed that the transient reflectivity of the WSe$_2$ in contact with the GaAs is smaller, almost half, at the wavelengths of resonance of the free exciton in WSe$_2$ when compared to the intensity of the hBN isolated region, indicating a higher absorption of the TMD in direct contact.
This response can be related to the change of the Fermi level due to the formation of the heterojunction, which reduces the electron density in the TMD.
Moreover, the charge transfer at the junction allows for the presence of free states in the WSe$_2$ conduction band, which can be accessed by the photo-excited electrons, thereby enhancing the absorption of the region.
In contrast, in the hBN-isolated area, stimulated emission and photobleaching will play an important role in reducing the absorption of the flake and increasing the reflectivity.
For the wavelengths in resonance with the free excitons in GaAs (800 nm - 830 nm), we observe a higher, negative, reflectivity in the sample in direct contact when compared to the isolated one.
This observation is consistent with an increment of the photoinduced absorption of GaAs, produced by the larger density of electrons in the substrate resulting from a shift of the bands in the heterojunction.

\begin{figure}[t!]
    \centering
    \includegraphics[width=\linewidth]{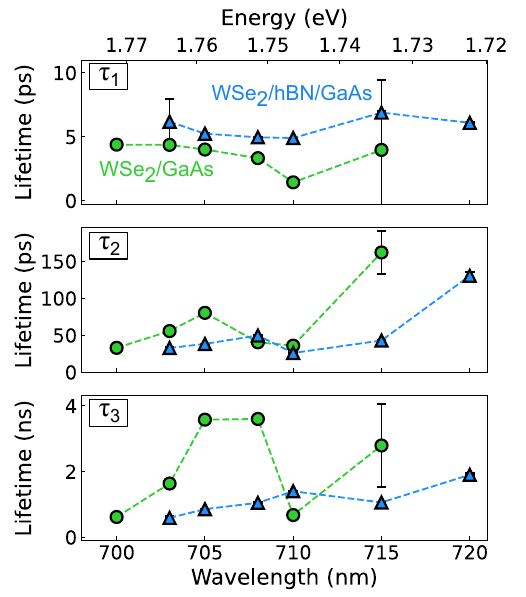}
    \caption{Lifetimes of the TRDR signals as a function of the wavelength extracted from the three exponential decay processes described in the main text. When not shown, the error bars, obtained by the fit, are smaller than the point size.}
    \label{Fig3:lifetimes}
\end{figure}

\begin{figure}[ht]
    \centering
	\includegraphics[width=\linewidth]{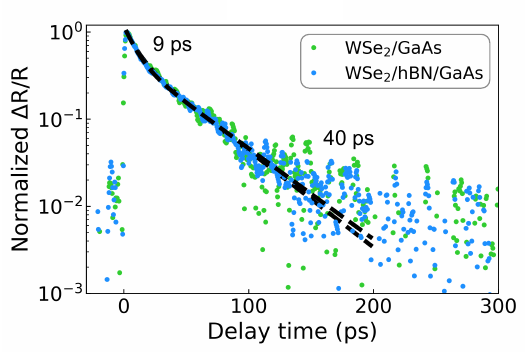}
	\caption{Normalized TRDR for the WSe$_2$/GaAs (green) and WSe$_2$/hBN/GaAs (blue) regions at room temperature. The laser was tuned to be resonant with the WSe$_2$ exciton, at $\lambda = 740$ nm.}
    \label{Fig4:RoomT}
\end{figure}

By fitting the TRDR of the measurements for different wavelengths, we extract the energy-dependence of the decay lifetimes of the two regions of interest, which are presented in Figure \ref{Fig3:lifetimes}.
We did not observe any clear trend with the wavelength for the fast decay ($\tau_1$) other than a slightly faster decay in the sample in direct contact as described previously.
On the other hand, the results for the second decay time ($\tau_2$) present a clearer trend, revealing one maximum lifetime at 705 nm for the TMD in direct contact with the GaAs substrate and at 708 nm in the isolated region, which matches with the resonances of WSe$_2$ exciton recombination of each area.
Furthermore, we observe another maximum, and the highest $\tau_2$ value, when exciting with a wavelength of 715 nm, which is related to the signal coming from the negative, less intense, differential reflectivity in Figure \ref{Fig2:wavelength}.b.
Data points for 720 nm in the direct contact region were discarded as the signal-to-noise ratio was too low to allow fitting.
Although negative signals are commonly associated with photoinduced absorption, it has also been observed that in TMDs, bandgap renormalization plays an important role in this effect \cite{Pogna2016}.
Therefore, we associate the different lifetimes obtained at this wavelength with the different origins of the relaxation path involved.
Lastly, the obtained long lifetimes ($\tau_3$) make clear the longer-lived character of the photo-excited carriers in the TMD in direct contact with the GaAs, close to the resonance. 

To determine the role of thermal effects, we measured the TRDR at room temperature (Figure \ref{Fig4:RoomT}), in the two regions of interest, exciting at the WSe$_2$ exciton resonance, $\lambda =$ 740 nm.
Our results show an overall shorter lifetime of the generated excitations when compared with the measurements at low temperature.
We observe a similar behavior for both regions of the sample with just two clear decay processes: a fast decay of 9 ps and a slower decay of around 40 ps.
If compared with our previous analysis at low temperature, we obtain a longer decay time $\tau_1$, a shorter decay time $\tau_2$ and the total absence of the presence of $\tau_3$ decay process at room temperature. 
These findings are in agreement with earlier studies reporting longer decay times $\tau_1$ of monolayer WSe$_2$ when increasing temperature \cite{Tengfei2014, Akmaev2020}.
One possible explanation for this phenomenon is the important role of dark states in tungsten-based TMD monolayers which are observed for instance in the enhancement of the photoluminescence when increasing the temperature \cite{Xiao2015}. 
In our experiments at low temperature, the excitation in resonance with low laser power results in a reduced source for the electrons in dark states to transit into bright states.  
At high temperatures, electron-phonon interactions mediate the transition and cause an increase in the population and lifetimes of the fastest process $\tau_1$ in both regions of the sample.
At the same time, this relaxation path, as well as other intralayer processes, becomes preferred over the charge transfer to the substrate, effectively eliminating at high temperatures the long-lived component $\tau_3$ of the dynamics which is observed at low temperatures.
Another possible relaxation channel is a change in the band alignment with the temperature. 
In this case, the small difference in the valence band maximum considered in Figure \ref{Fig1:summary}c could be enough for them to switch positions with the change in the temperature.
Under this hypothesis, WSe$_2$/GaAs would have a type-I band offset at room temperature and switch to a type-II band alignment when reducing the temperature.


Our observation of a type-II band alignment and charge transfer between the prototypical 2D/3D semiconductors, WSe$_2$ and GaAs, indicates the promise of using such junctions in future optical and optoelectronic devices \cite{Lin2015, Xu2016, Sarkar2020, Jia2019}.
The long-lived (3.5 ns) opto-excited carriers observed here should allow for a long enough time for these carriers to be transported away from the junction region, and used in photovoltaic devices.
Additionally, a long decay time is a crucial element for lasers.
Therefore, the combination of a long carrier lifetime with the unique spintronic properties of both WSe$_2$ and GaAs such as long spin lifetimes and electric control over the spin information \cite{Zutic2004, xu2014spin, salis2001electrical, gong2013magnetoelectric}, makes these junctions particularly appealing for lasers which make use of the spin degree-of-freedom, i.e. spin lasers \cite{vzutic2020spin}, which have been shown to be able to operate at much higher modulation frequencies than conventional lasers \cite{lindemann2019ultrafast}.
We envision that such junctions, as the one shown here, using novel 2D semiconductors in combination with well-established and industrially-proved 3D systems, can lead to an easier uptake of 2D materials in industrial settings, leading to new device architectures.


We thank J. G. Holstein, H. de Vries, F. van der Velde, H. Adema, and A. Joshua for their technical support.
This work was supported by the Dutch Research Council (NWO — STU.019.014), the Zernike Institute for Advanced Materials, and the Brazilian funding agencies CNPq, FAPEMIG and the Coordenação de Aperfeiçoamento de Pessoal de Nível Superior - Brasil (CAPES) - Project code 88887.476316/2020-00.
Sample fabrication was performed using NanoLabNL facilities.




\nocite{*}


\input{ref.bbl}

\end{document}

%% file: ref.bbl
%